\documentclass[aps,twocolumn,superscriptaddress,prl,showpacs]{revtex4}
\usepackage{amsmath}
\usepackage{graphicx}
\usepackage[usenames]{color}

\usepackage{bm}

\begin{document}
\title{Optically probing the fine structure of a single Mn atom in an InAs quantum dot}

\author{A.~Kudelski}
\author{A.~Lema\^{i}tre}\email{Aristide.Lemaitre@lpn.cnrs.fr, }
\author{A.~Miard}
\author{P.~Voisin }
\affiliation{Laboratoire de Photonique et Nanostructures-CNRS, Route de Nozay, 91460
Marcoussis, France}
\author{T.~C.~M.~Graham}
\author{R.~J.~Warburton}
\affiliation{School of Engineering and Physical Sciences, Heriot-Watt University, Edinburgh
EH14 4AS, United Kingdom}
\author{O.~Krebs}\email{Olivier.Krebs@lpn.cnrs.fr}
\affiliation{Laboratoire de Photonique et Nanostructures-CNRS, Route de Nozay, 91460
Marcoussis, France}

\def\xp{$X^{+}$}
\def\xm{$X^{-}$}
\def\x0{$X^{0}$}
\def\xx0{$2X^{0}$}
\def\etal{~\textit{et al.}~}
\date{\today}

\begin{abstract}

We report on the optical spectroscopy of a single InAs/GaAs quantum dot (QD)  doped with
a single Mn atom in a longitudinal magnetic field of a few Tesla. Our findings show that
the Mn impurity is a neutral acceptor state $A^0$ whose effective spin  $J\!=\!1$ is
significantly perturbed by the QD potential and its associated  strain field. The spin
interaction with photo-carriers injected in the quantum dot is shown to be ferromagnetic
for holes, with an effective coupling constant of a few hundreds of $\mu$eV, but
vanishingly small for electrons.
\end{abstract}

\maketitle

\begin{figure}[h]
\includegraphics[width=0.48 \textwidth,angle=0]{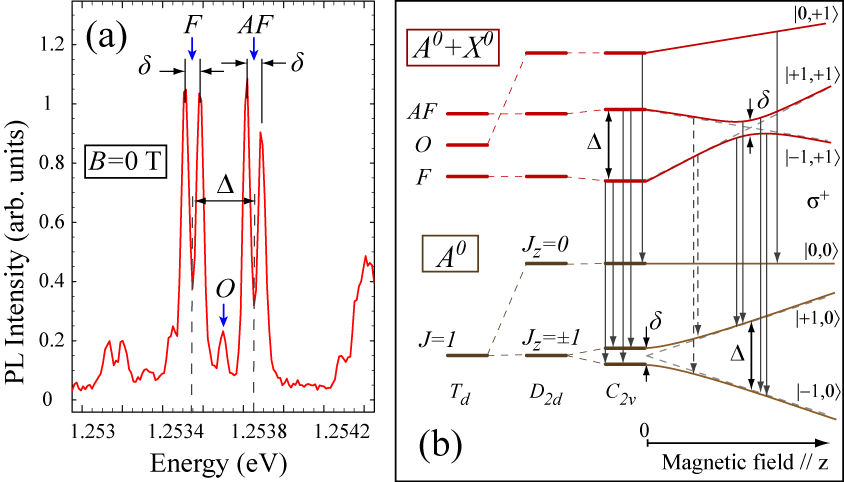}
\caption{(Color Online) (a) Micro-PL spectrum of an individual InAs quantum doped with a single
Mn atom in zero magnetic field at $T=2$~K. These lines were identified as originating from a
charged exciton $X^-$. (b) Schematic of the excitonic transitions from $A^0 + X^0$ (initial
state) to $A^0$ (final state) taking into account the local potential anisotropy and a
longitudinal magnetic field. For simplicity only the bright states corresponding to $\sigma^+$
polarized transitions  are shown. The same diagram holds for  $A^0 + X^-$ (see text) by
replacing the above labels by  $|J_z,+3/2\rangle$ ($|J_z,+1/2\rangle$) in the initial (final)
states respectively.} \label{fig1_Trion_B=0}
\end{figure}

\indent The spin state of a single magnetic impurity could be envisaged as a primary building
block of a nanoscopic spin-based device \cite{Efros01,Fernandez07} in particular for the
realization of quantum bits \cite{Leuenberger01}. However probing and manipulating such a system
require extremely high sensitivity.  Several techniques have been successfully developed over
the last few years to address a single or few coupled spins: electrical detection~\cite{Xiao04,
Elzerman04}, scanning tunneling microscopy (STM) \cite{Manassen89, Heinrich04, Hirjibehedin06,
Kitchen06}, magnetic resonance force microscopic \cite{Rugar04}, optical spectroscopy
\cite{Gruber97}. Recently, Besombes\etal\cite{Besombes04, Besombes05, Leger05a, Leger06} have
investigated the spin state of a single Mn$^{+2}$ ion embedded in a single II-VI self-assembled
quantum dot (QD). In this system the magnetic impurity is an isoelectronic center in a $3d^5$
configuration with spin $S=5/2$. The large exchange interaction between the spin of the
photocreated carriers confined inside the dot and the Mn magnetic moment induces strong
modifications of the QD photoluminescence (PL) spectrum: $2S+1=6$
discrete lines are observed, reflecting the Mn spin state at the instant when the exciton recombines. \\
 \indent The case of the Mn ion is different in GaAs, since the impurity is an acceptor
in this matrix with a rather large activation energy (113~meV). Two types of Mn centers exist in
GaAs, the $A^0$ and the $A^-$ states. In low doped GaAs (below $10^{19}$~cm$^{-2}$), the former
is dominant. It corresponds to the $3d^5+h$ configuration, where $h$ is a hole bound to the Mn
ion with a Bohr radius around 1~nm~\cite{Schneider87}.
 When considering a single Mn impurity in  InAs QD several issues arise: the
impurity configuration, its possible change when photo-carriers are captured, the
influence on the binding energy of excitonic complexes,  the strength and sign of the
effective exchange interaction with each of the carriers (electron or hole) in the QD
$S$-shell. In this Letter, we report the first evidences of a single Mn impurity in an
individual InAs QD which enable us to answer  most of the above questions. In particular,
we find that the formation of excitons, biexciton and trions is weakly perturbed by the
impurity center, whereas the effective exchange coupling with the Mn impurity (found in
the  $A^0$ configuration) is
ferromagnetic for  holes (a few 100~$\mu$eV's) and almost zero for the electrons.\\
 \indent The sample was grown by molecular beam epitaxy on
a semi-insulating GaAs [001] substrate. The Mn-doped quantum layer was embedded inbetween
an electron reservoir and a Schottky gate. This design gave us the possibility to observe
both neutral and charged excitons. It consists of a 200~nm thick $n$-doped GaAs layer
($n=2\times10^{18}$~cm$^{-2}$) followed by a non-intentionally doped (n-i-d) 20~nm GaAs
layer, the Mn-doped  QD layer, and capped with a n-i-d GaAs (30~nm)/
Ga$_{0.7}$Al$_{0.3}$As (100~nm)/ GaAs (20~nm) structure. The QD layer was formed by the
deposition of 1.7~ML of InAs during 5~s.
 The substrate temperature was set to
500~$^\circ$C (optimal for QD) during the growth of the whole structure. The Mn doping
was carried out by opening the Mn cell shutter during the QD growth. The cell temperature
was set to 590~$^\circ$C. The precise determination of the Mn atom density is difficult
in this material because of the large segregation of Mn atoms at these growth
temperatures as observed by STM [J.-C.~Girard, unpublished]. Estimations from Hall effect
measurements in thick and uniformly Mn doped GaAs layers grown at the same temperature
yielded a density of approximatively $1-2\times10^{11}$ Mn atoms per cm$^2$, giving a
probability of $1/3-2/3$ Mn per dot. However, in $\mu$-PL measurements on a large
collection of single QDs we observed only rare occurrences of Mn doping ($<$0.1\%),
likely due to the Mn segregation away from the QD layer. Samples grown  at higher Mn cell
temperature (660~$^\circ $C)
showed a much larger probability ($\sim$1\%)  of finding dots containing a single Mn atom.\\
\indent The $\mu$-PL spectroscopy of individual InAs:Mn QDs was carried out with a split-coil
magneto-optic cryostat. A 2~mm focal length aspheric lens (N.A.~0.5) was used to focus the
He-Ne excitation beam and to collect the PL from the sample, while the relative positioning in
all three directions  was ensured by Attocube$^{TM}$ piezo-motors. 
 All measurements presented in this Letter were performed at low
temperature (T=2~K) and the magnetic field was applied parallel to the optical axis (Faraday
configuration). The PL was dispersed by a 0.6~m-focal length
double spectrometer and detected by a Nitrogen-cooled CCD array camera.\\
\indent  We first present the optical signature in zero magnetic field  of a single Mn atom
uncovered by our experiments in  the PL spectrum of about 10 different QDs. A characteristic
spectrum is shown in Fig.~\ref{fig1_Trion_B=0}(a). It consists of two bright doublets labelled
$F$ and $AF$ separated by an energy $\Delta$ of the order of a few 100~$\mu$eV, and of a weaker
central line denoted by $O$. The splitting $\delta$ of the doublets is the same for both lines
$F$ and $AF$ and typically amounts to a few tens of $\mu$eV.
A simple  interpretation of this spectral feature can be constructed by assuming that the QD
contains a Mn impurity in the $A^0$ configuration, i.e. a hole bound to an $A^-$
center~\cite{Schneider87,Bhattacharjee99,Govorov04}. This acceptor state is characterized in
bulk GaAs  by an anti-ferromagnetic ``$p$-$d$'' exchange between the 3$d^5$ Mn spin $S=5/2$ and
the hole spin $J_h=3/2$. The latter takes the form of a Heisenberg Hamiltonian $\varepsilon\,
\bm{\hat{S}}\cdot\bm{\hat{J}_h}$~\cite{Bhattacharjee99} (with $\varepsilon \sim$5~meV), giving
rise to splittings of  $A^0$ eigenstates as a function of the total angular momentum
$\bm{\hat{J}}=\bm{\hat{J}_h}+\bm{\hat{S}}$ : the triplet ground state $J\!=\!1$
($J_z\!=\!\pm1,0$) turns out to be well separated from the higher levels $J\!=\!2,3,4$ by at
least $2\varepsilon\sim 10$~meV. Therefore, at low temperature when $k_B T\ll2\varepsilon$ the
Mn impurity is completely thermalized in its ground state. Assuming similar ``$p$-$d$'' exchange
in  InAs QDs, we may  consider that the photo-carriers captured by the QD interact with an
effective spin $J\!=\!1$. Such a situation is depicted  in Fig.~\ref{fig1_Trion_B=0}(b) in the
case of an electron-hole pair (or neutral exciton \x0) in its ground state. We focus here only
on the exciton bright states with projection of angular momentum $J^{exc.}_z\!=\!\pm1$, since
the dark states ($J^{exc.}_z=\pm2$)  do not contribute to the PL signal. A natural basis to
describe the $A^0+X^0$ initial states of the excitonic transition reads thus
$|J_z,J^{exc.}_z\rangle$. In case of exchange interaction between $A^0$ and \x0, these levels
are split into three doubly degenerate levels which read $|\pm 1,\pm 1\rangle$, $|0,\pm
1\rangle$, $|\mp 1,\pm 1\rangle$ corresponding to a ferromagnetic ($F$), ``orthogonal'' ($O$),
and anti-ferromagnetic ($AF$) spin configuration, respectively. If the final $A^0$ states were
perfectly degenerate, as predicted for the acceptor level in bulk GaAs, then we should observe
in the PL spectrum three lines equally spaced and of identical intensity, similar to the six
lines observed in CdTe QDs doped with a single Mn atom~\cite{Besombes04}. Actually, due to its
hole component,  the $A^0$ state is sensitive to  local variations of composition and strain
over a typical distance of 1~nm from the impurity center~\cite{Schneider87,Govorov04,Yakunin07}.
In self-assembled InAs QDs which ressemble a flat lens of $\sim$4~nm height, the most important
perturbation of the bulk potential occurs along the growth direction, $z$. Such a perturbation
of $D_{2d}$ symmetry shifts the  $J_z=0$ level to higher energy with respect to the
$J_z=\pm1$ states. 
 The same effect occurs for the $O$ level in the $A^0+X^0$
complex as this perturbation is diagonal in the basis we have chosen. If in addition the
potential experienced by the impurity has some in-plane anisotropy (with $C_{2v}$
symmetry or lower), then the $J_z=\pm1$ are further split by an energy $\delta$. Such an
effect is expected because the Mn impurity is very likely not in the center of the QD.
Note that this anisotropy acts perturbatively as an off-diagonal term for the $A^0+X^0$
levels which are already split by $\Delta$. Following this scheme, the optical
transitions from the $F$ and $AF$ levels appear as doublets due to the \emph{final state}
splitting, whereas the $O$ level may recombine only to the $J_z=0$ state because of
orthogonality of its $A^0$ component ($J_z=0$) with the $J_z=\pm1$ states. The shift of
the $O$ level does not reflect in the transition energy, since it appears in both the
initial and final states,
 however it explains the  weaker intensity observed experimentally for the $O$ line because of $A^0$ thermalization
 in the $J_z\!=\!\pm1$ levels.\\
\indent The main support to this interpretation comes from the evolution in a longitudinal
magnetic field. Thanks to the Zeeman effect it is possible to  restore the $A^0$ eigenstates to
$J_z\!=\!\pm1$. The $F$ and $AF$ doublets should  transform to single lines for a field
$B_z>\delta/2 g_{1}\mu_\text{B}$ where $g_{1}$ is the $A^0$ $g$~factor in the $J\!=\!1$ spin
configuration and $\mu_\text{B}$ is the Bohr magneton. Taking the value $g_{1}=2.77$ found for
GaAs:Mn~\cite{Schneider87} the typical magnetic field required amounts  to only 230~mT for the
QD shown in Fig.~\ref{fig1_Trion_B=0}. In parallel, the magnetic field splits the $F$ and $AF$
levels by the sum of Zeeman effects for $A^0$ and $X^0$. Therefore, the Zeeman splitting of
$A^0$ does not reflect straightforwardly in the PL spectra  apart for  the ``forbidden''
transitions involving a spin-flip of $A^0$ and represented by dashed arrows in
Fig.~\ref{fig1_Trion_B=0}(b). When the magnetic field reaches the value $\Delta/2
g_{1}\mu_\text{B}$ ($\sim$1~T in our case) the $|+1,+1\rangle$ and $|-1,+1\rangle$
 states are now brought into coincidence. Since they are
formed with the same exciton spin, the anisotropic interaction between the $J_z=\pm1$
levels splits the $A^0+X^0$ levels by the same energy splitting $\delta$ as in zero
field. For this very specific field the PL spectrum should thus be quite similar to the
spectrum in zero field as illustrated in Fig.~\ref{fig1_Trion_B=0}(b), with the
splitting $\Delta$ ($\delta$) in  the final (initial) states.\\
\begin{figure}[h]
\includegraphics[width=0.4 \textwidth,angle=0]{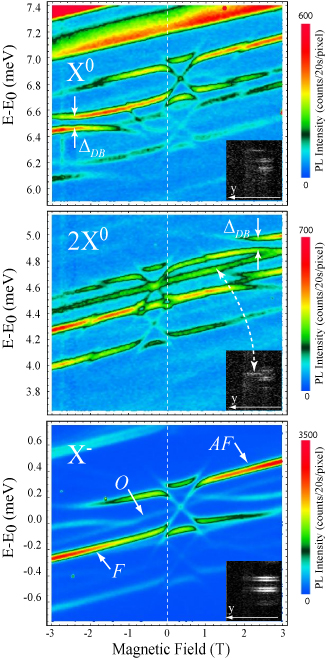}
\caption{(Color online) Contour-plot of $\mu$-PL intensity from a single InAs QD doped with a
single Mn impurity against longitudinal magnetic field and detection energy $E-E_0$
($E_0=1.2536$~eV). The three energy windows  correspond to the  \x0, \xx0 and \xm$\,$ states
 observed in the same $\mu$-PL spectra. Insets : CCD images (100 pixels $\times\sim$~150 pixels) at
$B_z=0$~T showing the spatial correlations of the lines along the vertical axis $y$ of the
set-up. An extra line coming from another QD (dashed arrow) is superposed on  the \xx0
feature.} \label{Fig2}
\end{figure} \indent To study the magnetic field dependence of the $X^0$-to-$A^0$ coupling, we recorded a
series of 121 $\mu$-PL spectra over a 10~meV-energy range, by varying the magnetic field
from $-3$~T to $+3$~T with a step of 50~mT. The detection was set $\sigma^+$ to help
identify the different levels and their interactions.  The  $\mu$-PL intensity was
plotted on a color-scale against magnetic field and energy-detection, using an
interpolating function for graphical rendering. To focus on the spin-dependent
interactions we subtracted the diamagnetic-shift $\propto B_z^2$. Figure~\ref{Fig2}
displays three spectral regions of this contour-plot, showing clearly correlated spectral
lines that could be identified (after a careful analysis) as the three excitonic features
$X^0$, $2X^0$ and $X^-$ originating all from the same individual QD. Remarkably, the
$2X^0$ and $X^-$ set of lines are separated from $X^0$ by  roughly the same  binding
energies than in undoped InAs QDs emitting at $\sim$1.25~eV~\cite{PRB-Eble}. Note Fig.~\ref{fig1_Trion_B=0}(a) is the cross-section at $B=0$~T of the \xm contour-plot.\\
\begin{figure}[h]
\includegraphics[width=0.4 \textwidth,angle=0]{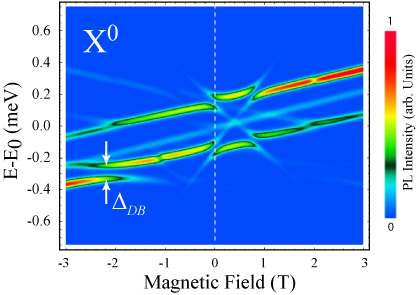}
\caption{(Color online) Theoretical contour-plot of \x0 PL intensity from a single InAs QD doped
with a single Mn impurity against longitudinal magnetic field and energy.} \label{Fig3}
\end{figure}
\indent The main feature common to the plots of Fig.~\ref{Fig2} is a very peculiar pattern
resulting from the evolution of the  zero-field doublets to another pair of doublets at
$B\approx 0.75$~T. The resulting crossing lines correspond to the ``forbidden'' transitions
involving $A^0$ spin-flip from $J_z=\pm1$ to $J_z=\mp1$ respectively. Obviously these
transitions are not strictly forbidden because of the anisotropic coupling either in the final
state (at $B=$0), or in the initial state (at $B\approx$0.75~T). 
 Focusing on the \xm$\,$
feature, we clearly observe a strong evolution of the intensity ratio between the $F$ and $AF$
lines due to the $A^0$ thermalization on one of the $J_z=\pm1$ levels depending on the field
direction~\footnote{Thermalization effects are assumed negligible in the excitonic states due to
a long spin lifetime compared to the recombination time. See also Ref.~\cite{Besombes04}.}. For
$B_z>0$, the $|+1,+1\rangle$ ($|-1,+1\rangle$) population should decrease (increase).
 Actually, it is this simple feature which
allowed us to ascribe confidently the low energy doublet $F$ to the ferromagnetic
$A^0+X^0$ configuration. We note that such an effective ferromagnetic coupling was
already reported by J.~Szczytko\etal\cite{Szczytko96} in very
dilute Ga$_\text{1-x}$Mn$_\text{x}$As ($x<0.001$).\\
\indent In each  case shown in Fig.~\ref{Fig2}, an exact replica of the main pattern is found at
lower energy. We  ascribe them to temporal electrostatic fluctuations of the QD environment
which rigidly shift all the excitonic lines, e.g. due to  charge  trapping and detrapping  in
the QD vicinity. Since these replica were not found for other Mn-doped QDs that we have examined
(and showing also the same cross-like patterns) we conclude that they are not related to the
intrinsic signature of a Mn-impurity. We chose to show
this particular dot because three excitonic complexes were simultaneously visible with a  high signal to noise ratio.\\
\indent Another striking feature is the symmetry
between \x0 and \xx0. 
 It results from the polarization correlation in the biexciton cascade imposed by the
Pauli principle. As we detect only $\sigma^+$ photons the measured transitions from \xx0 lead
to the $\sigma^-$~polarized~\x0, which obviously has the same field dependence as the
$\sigma^+$~polarized~\x0 but for $B_z\rightarrow -B_z$.
This observation strongly supports the line identification and  actually  indicates that
the biexciton (with both holes and electrons in singlet spin configuration) has no spin
interaction with the Mn impurity. Note that the very same symmetry has been observed in Mn-doped CdTe QDs~\cite{Besombes05}.\\
\indent Finally, the position of the cross-like pattern for the  \xm$\,$case is very
instructive. It reveals  that one of the electron-$A^0$ or hole-$A^0$ exchange integrals
must be vanishingly small with respect to the other. If not the  mixing at $B_z=0$
between the $J_z\!=\!\pm 1$ states would be reduced both in the initial state (due to
hole-$A^0$ exchange) and final state (due to electron-$A^0$ exchange). There would be no
splitting  and the cross-like pattern would be shifted to a different field. Since it
appears at the same positive field as for \x0, the \xm$\,$transitions  must be described
by the  diagram of Fig.~\ref{fig1_Trion_B=0}(b), yet with  $e$-$A^0$ as the final state.
We can therefore conclude that the electron-$A^0$ coupling  is negligible as compared to
$\delta$ (actually below 20~$\mu$eV from a precise comparison of the spectra at $B_z=0$).\\
 \indent To support  the  above discussion,  we have modeled the spin interactions with the
 Mn impurity for the three excitonic configurations. To reproduce all details of our
 experimental results, it appeared necessary to include not only the $J\!=\!1$ states of
 $A^0$ but also the $J=2$ states. Our model includes  the Zeeman Hamiltonian for a single
 particle (Mn, bound hole $h_1$, QD $S$-shell hole $h_2$ and  electron $e$), strain
 Hamiltonian for $h_1$~\cite{Yakunin07}, valence band mixing between light- and
 heavy-components for $h_2$~\cite{Kowalik07}, and
  exchange interaction within each pair of particles. A detailed
 discussion of this model will be published elsewhere. We present in
 Fig.~\ref{Fig3} the contour plot of theoretical PL spectra corresponding to the  \x0-$A^0$
 configuration. By adjusting strain and exchange parameters, our model reproduces remarkably well the cross-like
 pattern, the effect of  Mn thermalization ($T_\text{Mn}$=10~K) as well as the anticrossing
 $\Delta_{DB}$  observed at -2~T. The latter results from a coupling between the bright $F$
 exciton $|+1,+1\rangle$ and the dark $AF$ exciton $|+1,-2\rangle$ when they are brought into
 coincidence by the field. Our model reveals that this is a resonant third order coupling
 involving  the  $h_2$ valence band mixing, a shear strain $\epsilon_{xz}$ (which also
 contributes to $\delta$) and the effective $h_2$-$A^0$  exchange constant  $\Delta_{12}$
 between  the $A^0$  spin subspaces $J=1$ and $J=2$. It reads
 $\Delta_{12}=\varepsilon_{h_2\text{-Mn}}-\varepsilon_{h_1\text{-}h_2}$ where $
 \varepsilon_{\alpha\text{-}\beta}$ is the exchange integral between the spins $J_{\alpha}$
 and $J_\beta$.  To reproduce our experimental results we found that the $F$-$AF$
 splitting $\Delta$ is  dominated by this  exchange term $\Delta_{12}$ while the exchange
 term $\Delta_{11}=(7\varepsilon_{h_2\text{-Mn}}-3\varepsilon_{h_1\text{-}h_2})/4$ within the
 $J=1$  subspace contributes less than 10\% of $\Delta$.\\

 \indent In conclusion, the successful
$\mu$-PL investigation in a longitudinal magnetic field of a single Mn-doped InAs quantum
dots reveals remarkable features bringing new insights into the spin interactions between
carriers and a Mn impurity in a III-V matrix. The anti-ferromagnetic coupling between the
hole bound to the magnetic impurity and the $3d^5$ Mn electrons is confirmed. In
contrast, the effective coupling of the Mn impurity as a whole  ($A^- + h$) with a hole
confined in an InAs QD is proven to be ferromagnetic, while it essentially vanishes for a
confined electron. The influence of the strain field on the Mn acceptor level is clearly
evidenced, and gives rise to a very specific spectral signature of the Mn doping. Our
results reveal that the Mn spin in $A^0$ configuration  represents a  two-level system
well separated from higher energy levels which opens new outlooks  for spin-based quantum
information processing,
e.g. by exploiting  the exchange interaction with optically polarized carriers.\\
\begin{acknowledgments}
This work was partly supported by the European Network of Excellence SANDIE, the ANR
contracts BOITQUANT and MOMES.
\end{acknowledgments}
\bibliographystyle{apsrev}

\end{document}